\documentclass{IEEEcsmag}

\usepackage[colorlinks,urlcolor=blue,linkcolor=blue,citecolor=blue]{hyperref}

\newcommand{\pra}{Phys. Rev. A}   

\usepackage{cite}
\usepackage{amsmath,amssymb,amsfonts}
\usepackage{algorithmic}
\usepackage{graphicx}
\usepackage{textcomp}
\usepackage{xcolor}
\def\BibTeX{{\rm B\kern-.05em{\sc i\kern-.025em b}\kern-.08em
    T\kern-.1667em\lower.7ex\hbox{E}\kern-.125emX}}
\usepackage{upmath}

\jvol{XX}
\jnum{XX}
\paper{8}
\jmonth{}
\jname{ }
\pubyear{}

\begin{document}


\title{A Portal for High-Precision Atomic Data and Computation: Design and Best Practices}

\author{}
\affil{\textit{This work has been submitted to the IEEE for possible publication. Copyright may be transferred without notice, after which this version may no longer be accessible.}}

\author{Parinaz Barakhshan}
\affil{University of Delaware, Department of Electrical and Computer Engineering}

\author{Akshay Bhosale }
\affil{University of Delaware, Department of Electrical and Computer Engineering}

\author{Amani Kiruga }
\affil{University of Delaware, Department of Computer and
Information Sciences}

\author{Rudolf Eigenmann}
\affil{University of Delaware, Department of Electrical and Computer Engineering}

\author{Marianna S. Safronova}
\affil{University of Delaware, Department of Physics and Astronomy}

\author{Bindiya Arora}
\affil{Perimeter Institute for Theoretical Physics and Guru Nanak Dev University }

\markboth{Department Head}{Paper title}

\begin{abstract}
The \textit{atom portal}~\cite{UDportal}, udel.edu/atom, provides the scientific community with easily accessible high-quality data about properties of atoms and ions, such as energies, transition matrix elements, transition rates, radiative lifetimes, branching ratios, polarizabilities, and hyperfine constants. The data are calculated using a high-precision state-of-the-art linearized coupled-cluster method, high-precision experimental values are used where available. All values include estimated uncertainties. Where available, experimental results are provided with references.
This paper provides an overview of the portal and describes the design as well as applied software engineering practices.
\end{abstract}

\maketitle
\chapterinitial{Introduction} 
\label{sec:Intro}
In a number of present applications, ranging from studies of fundamental interactions to the development of future technologies, accurate atomic theory is indispensable to the design and interpretation of experiments, with direct experimental measurement of relevant parameters being impossible or infeasible. These data are also in high demand by broader atomic, plasma, astrophysics, and nuclear physics communities. The need for high-precision atomic modeling has increased significantly in recent years with the development of atomic-based quantum technologies for a wide range of fundamental and practical applications. Further rapid advances will require accurate knowledge of basic atomic properties, most of which remain highly uncertain and difficult to measure experimentally. The lack of a comprehensive atomic database for modern research applications created a bottleneck in experimental design and analysis as well as industry applications which motivated the creation of the portal for high-precision atomic data and computation.

This paper reports the release of Version 2.0 of an online portal that provides high-precision atomic data derived from computationally intensive calculations performed on a growing number of atoms and ions. This project fills the need for high-quality atomic data and software in several scientific communities. 
The latest version of the portal, Version 2.0, provides over half a dozen properties of atoms and ions, including highly charged ions.

As of November 2022, over 2600 users have accessed the website since it was published in April 2021. Users are connecting from 93 countries. The top ten countries that have accessed the website were the United States (36\%), Germany (11\%), India (10\%), United Kingdom (6\%), China (5\%), Canada (3\%), Switzerland (3\%), Singapore (3\%), France (2\%), and Russia (2\%). There have been  5,567 downloads and prints from the website.

The latest version of the portal has added new properties and elements as well as enhanced the user interface based on community feedback.
The majority of web pages are generated in an automated manner from the data supplied by the physicists. In this way, human error and repetitive work for different elements and properties can be reduced and the creation of pages for newly introduced elements is made easy for non-experts.

The project is an interdisciplinary effort where physicists have produced the data and computer scientists have developed the website for presenting the data. Software engineering approaches have been applied to facilitate this joint portal  development, resulting in savings of time and effort.

The remainder of the paper is organized as follows. 
The \textbf{portal overview} section provides a brief overview of the portal. A description of how the data is generated is provided in the \textbf{generation of the Physics data} section. The section on \textbf{portal design challenges} describes the challenges we encountered during the development of the portal. The practices employed in this project are described in the section entitled \textbf{applied software engineering practices}, followed by \textbf{conclusions}.

\section{Portal Overview}
The current version of the portal provides users with pre-computed data on atomic properties through an interactive interface, allowing them to view and download the provided information. This information includes the properties -- \textit{transition matrix elements, transition rates, radiative lifetimes, branching ratios, hyperfine constants, quadrupole moments, scalar and dynamic polarizabilities}, and \textit{energies} of atoms and ions, including Li, Be$^{+}$, Na, Mg$^{+}$, K, Ca$^{+}$, Rb, Sr$^{+}$, Cs, Ba$^{+}$, Fr, and Ra$^{+}$. Furthermore, 13 highly charged ions have been added to this version, including Cs$^{6+}$, Ba$^{7+}$, Ce$^{9+}$, Pr$^{10+}$, Nd$^{11+}$, Nd$^{12+}$, Nd$^{13+}$, Sm$^{13+}$, Sm$^{14+}$, Sm$^{15+}$, Eu$^{14+}$, Cf$^{15+}$, and Cf$^{17+}$. 

\begin{figure}[h]
\centering
\includegraphics[scale=0.37]{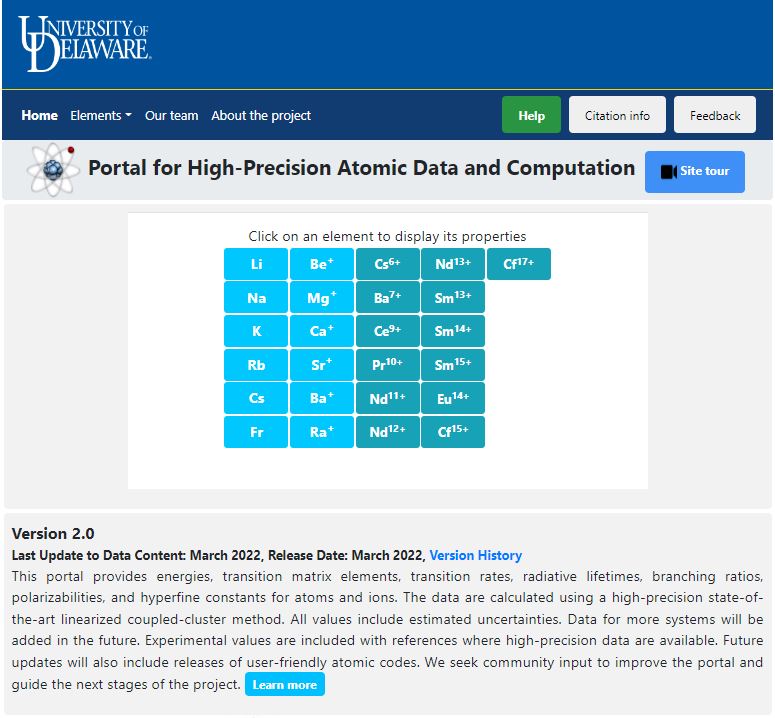}
\caption{Home page of the \textit{atom portal} at udel.edu/atom}
\label{fig:front}
\end{figure}

Figure \ref{fig:front} 
illustrates the home page of the Atom portal, which provides instant access to information about atoms and ions based on pre-computed properties.

\begin{figure}[h]
\centering
\includegraphics[scale=0.37]{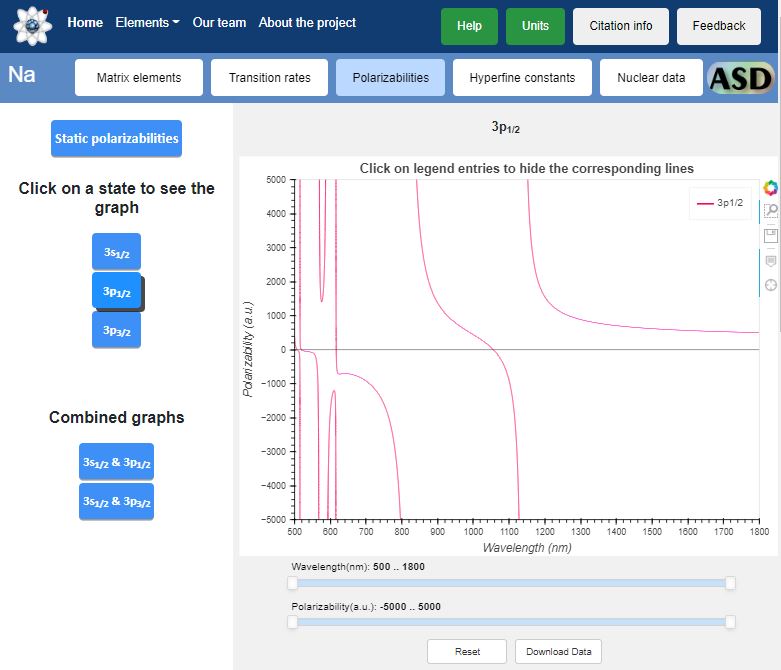}
\caption{Display of  \textit{polarizability} for element \textit{Na}}
\label{fig:pol}
\end{figure}

An example property calculated for these atoms is their polarizability, which describes their response to an external electric field. Figure \ref{fig:pol} 
illustrates the graphical interface designed for easy interaction by the user.

Data for 12 elements are calculated using a state-of-the-art linearized coupled-cluster method that will be discussed in more detail in the next section. High-precision experimental values are used where available with references. Data for highly-charged ions is taken from the literature and references to the original source. The goal of the portal is to provide a recommended value that users should use for any given property, i.e. this is the most accurate value given the state-of-the-art knowledge in atomic physics. 

\section{Generation of the Physics Data}

This section explains how physicists generate data. 
Data provided by the Atom portal for  Li, Be$^{+}$, Na, Mg$^{+}$, K, Ca$^{+}$, Rb, Sr$^{+}$, Cs, Ba$^{+}$, Fr, and Ra$^{+}$ are computed and evaluated for accuracy based on a long history of Physics research in developing accurate models for the computation of atomic properties and corresponding computational codes  using a linearized coupled-cluster method reviewed in \cite{2008Allorder}. The polarizability computations are described in \cite{2007Magic}. The computations of the quadrupole moments are described in \cite{2008Quadrupole}.

This method is also referred to as the all-order method in literature as it involves summing series of dominant many-body perturbation terms to all orders. In the single-double (SD) all-order approach, single and double excitations of the Dirac-Fock orbitals are included. The single, double, particle triple (SDpT) all-order approach also includes classes of the triple excitations. Omitted higher excitations are estimated by the scaling procedure which can be started from either SD or SDpT approximations. We carry out four all-order computations for each of the electric-dipole matrix elements, \textit{ab initio} SD and SDpT and scaled SD and SDpT. Either SD or SD scaled data are taken as final values based on the comparison of different contributions to the matrix elements.

An algorithm is used to determine the uncertainties of the electric-dipole matrix elements based on the spread of the four results, size of the correlation correction, and comparison of different contributions to the matrix elements~\cite{safronova2001high}.
A detailed description of the SD and SDpT all-order approaches with all formulas is given in the same document.

The computational codes have been made efficient through parallel computation using OpenMP pragmas embedded in the Fortran 90 code and running on multi-core machines.
Due to the high resource demand for the code runs, computations are performed on compute nodes of the Caviness computational cluster~\cite{CavinessIT} at the University of Delaware (UD).
These code runs calculate the broad range of atomic properties mentioned in the introduction.

Energies provide information about the electronic energy level relative to the ground state energy for each of the elements. The wavelength refers to photon wavelength for the transition between two electronic energy levels.  Transition matrix elements quantify the strength of electronic transition between two levels. Transition rates give the probability of an electron making the transition. Radiative lifetimes are the average times that an excited atom can remain in that state before emitting a photon. Branching ratios describe the relative probabilities of all possible transitions.  Hyperfine constants describe the interactions between the nucleus and the electrons and are used to quantify the resulting splitting of energy levels due to this effect. Quadrupole moments give information about the shape of an atoms' electron cloud. The scalar and dynamic polarizabilities describe the atoms' response to an external electric field; dynamic polarizability is used to  understand the response of an atom to a laser light of a given frequency. 

The data for  Cs$^{6+}$, Ba$^{7+}$, Ce$^{9+}$, Pr$^{10+}$, Nd$^{11+}$, Nd$^{12+}$, Nd$^{13+}$, Sm$^{13+}$, Sm$^{14+}$, Sm$^{15+}$, Eu$^{14+}$, Cf$^{15+}$, and Cf$^{17+}$ highly-charged ions have been computed by the University of Delaware 
team and collaborators using various methods, including the all-order method described above, and other methods that combine configuration interaction and coupled cluster approaches \cite{CI}. The use of more complicated methods is required due to the complicated electronic structures of some of the ions evaluated. As these computations are recent and represent the state-of-the-art, we use the generated data for the portal and provide references to original papers. 
\section{Portal Design Challenges}

This section describes the challenges faced in creating a portal that is easy to populate with data for physicists and easy to navigate for the user community.

\subsection{Automated portal generation from physics data}
The portal interface  provides functions that allow the user to select the data, present it in a user-oriented way, and print/download the viewed information. Users can select a particular atom or ion, choose one or several electronic states, and obtain data for a particular property they want to view.  The original data underlying the portal is provided by physicists in well-formatted files that are placed in a
source directory. The names and format of all data files are standardized for automated input. 

One design goal was to avoid manual portal changes when physicists
provide or update new data. Data updates could stem from improvements of the computational codes or from high-precision physics experiments conducted by any researcher in the community. Another goal was to create the user interface functionality without the need for dynamic  web
content support, making the website portable to a large range of
potential web servers. We addressed these challenges through the
development of a portal-generation algorithm, implemented through Python
scripts. The algorithm produces static web pages with the physics data
directly embedded, and with JavaScript methods providing the needed
user functionality.

\subsection{Data interface for physics developers} 
We aimed to make it
easy for physicists, both those involved in the project and external
researchers providing experimental data, to supply their information
without needing to understand  web  technology. To this end,
we defined appropriate data formats for expressing elements, ions, and
their properties. The data is in part generated automatically by the
computational codes. Manual assistance is needed for such tasks as
ensuring data validity, providing data from physics experiments, and
uploading the data to the portal's source directory. The chosen format uses CSV (comma-separated values), which can also be generated from an excel spreadsheet. 

\subsection{Integrating data from different sources}
The provision of data from two sources -- the computational codes and from external experimental measurements -- poses challenges for both the user interface and the implementation. For the user interface, we
chose a display that marks data from sources other than our
computational codes with a 
\raisebox{-0.2\height}[0pt][0pt]{\includegraphics[width=0.030\textwidth, height=0.4cm]{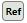}} 
tag. 
Clicking on this tag opens a dialog box providing a reference and a link to the paper describing the experiment. The presence of experimental data is signaled to the
portal-generation algorithm by the presence of a corresponding CSV
file. The algorithm reads this file and replaces the computed properties
by the experimentally provided values in the display, adding the tags.

\subsection{Presentation of complex state designations}
Physicists use complex
notations for electron states, which involve superscripts, subscripts,
and fractions. An example of such a complex notation is \(5f6p^2\;^2F_{5/2}\). In general, web page designers can use HTML notation for displaying such
expressions. Alternatively, LaTeX libraries support the same. The
state designations also pose challenges for the formats of physics
input and download files, which use plain text CSV. While there are
Unicode representations of superscripts and subscripts for plain text, many
viewing tools only support ASCII code; they display special Unicode
characters as ``unknown", such as question marks or blank squares. 
We discarded this solution, as we do not control the
users' viewer tools. The variance in the first part of the designation requires formatting decisions (consider $6s6p$, $6s^26p$, $6s^26p^2$, and $5d^36s6p$ all of which appear for some atoms). Our solution is to employ a plain-text, ASCII
representation of the states. We refer to this representation as the
{\em dot notation}. For example, in dot notation, the example state mentioned above is represented as 5f.6p2 2F5/2, which indicates there is one 5f electron and two 6f electrons in this electronic state designation, allowing for automated input and conversion to the HTML.  The second part of the designation always has a format of a superscript, an upper case letter, and a subscript, simplifying automation.  Such formatting allows to describe complicated state designations and is already in use by the physics community \cite{NIST}. An alternative
would be the use of a LaTeX math notation, which is also an ASCII. The LaTeX representation of the above example  would be~\verb+5f6p^2\;^2F^0_{5/2}+.

We preferred dot notation for its simplicity and the fact that it is already in use by physicists. Furthermore, for internal processing, the states are enumerated. A search function, for example, can look up the state's number, instead of performing a string search operation.

\subsection{Eliminated redundancy in menu navigation}
In our initial version of the portal, it was possible to navigate to certain information pages on more than one path. For example, to see the transition rates for atom K,
one could choose the Transition Rate tab on the front page and then
select the atom.  Alternatively, the user could choose K in the ``Transition Rates" menu, which was available on all pages. While convenient, this form of redundant path navigation
was confusing to some users. We eliminated the redundancy in Version~2.

\subsection{Links to  related external websites} 
The information provided by the Atom portal is complimentary to the data
provided by other websites. In atomic structure, the most commonly used database is NIST's ASD (Atomic Spectra Database) \cite{NIST}. It has the largest collection of measured atomic energy levels and transition rates. However, while the accuracy of experimental energies is higher, the accuracy of experimental transition rates is generally low, with a few exceptions for most commonly studied elements. The database only focuses on these properties. To avoid
redundancy, we use ASD energy data where available, creating references to the ASD website instead of duplicating ASD data. While redundant data could add convenience of having all data in the same page, for example -- users would like to see energy levels next to transition rates  -- pointing the user to the ASD website better acknowledges their contribution and improves maintainability, in case NIST updates ASD information. The transition rates on our website are expected to be more accurate than on the ASD website, and we do not use ASD data. We also provide energy levels that are omitted on the ASD website due to lack of experimental data. 

\subsection{Data consistency checks}
Ensuring that the data shown on the Atom portal is always correct is a non-trivial task. Currently, substantial manual work is involved in verifying correctness, while certain checks
are automated already. For example, the display format for uncertainty is provided in the physics input files and also checked by a method in our portal-generation algorithm. To eliminate the need for manual intervention, it is vital to automate the data consistency checks, and is being addressed in our ongoing work.

\subsection{User feedback function}
Websites can only reach maturity with substantial user feedback. A prominent user-feedback button serves to solicit suggestions for improvements from portal users. The same function allows users to see suggestions made by others, including the implementation status. Users can also view the features that are forthcoming in the immediate and future releases.

\subsection{Tracking Portal Use}
The Google Analytics tool allows website administrators to track and analyze users who visit their websites. A Google Analytics script has been embedded in each page to collect data, such as the number of users, session statistics, approximate geographical location, browser type, device information, and the number of downloads and prints. 
This information allows us to understand the portal's use and make data-driven design decisions.

\section{Applied Software Engineering Practices}
\label{sec:BestPractices}
Careful attention to software engineering principles not only increases maintainability and  sustainability; it is also a  requirement of our project sponsor, the National Science Foundation (NSF). We describe the practices, as defined in an effort to develop best practices for multidisciplinary projects, in which research software engineering teams support, and collaborate with, domain science groups~\cite{https://doi.org/10.1145/3491418.3530293}. We briefly explain each of the recommended practices, followed by the specific use in our project and illustrating examples.
The recommendations identify two categories:
(i) best practices for project collaborations and (ii) best practices for software development. For reference, we use the same titles of the practices;
we also use the term Xperts for computer and data experts supporting domain scientists (physicists, in our project). 

\subsection{Diversity of Xpert Backgrounds}
Software development teams often include experts with diverse backgrounds. As a result, training should be planned so that less-familiar practices can be acquired as needed~\cite{ICS2019report}.

In the Atom project, developers from different backgrounds joined the development team at various stages of the development process. To be able to contribute to the project, each participant had to undergo a learning process.

For example, to automate the creation of web pages, developers from both computer science and physics departments were added to the project for a limited time. The on-boarding process included steps,  such as becoming familiar with the code, gaining the skills needed to make modifications, and asking questions.

\subsection{Breadth of Xpert Skills Needed}
Computational and Data-intensive (CDI) applications today incorporate a broad range of technologies, including computing paradigms, programming languages, architectures, and algorithms. 

The technologies to be used in the project and the skills required for the development team were determined from the project requirements. We then engaged project members with the necessary expertise.

\subsection{Collaborative Assistance Between Xperts and Domain scientists}
A recommended form of interaction between Xperts and domain scientists is through collaborative assistance, where Xperts work side by side (physically or virtually) with the domain scientists whose projects they support. Throughout the joint work, the collaborators tend to pick up sufficient knowledge of each others' skills and terminology. Supporting and facilitating collaboration between the members of the project is crucial. Joint, periodic  reviews of milestones will reduce issues that may impede progress. 

The project engaged both physicists and computer scientists in close collaboration. The collaboration involved weekly meetings to discuss the progress, provide updates, possibly revise decisions, assign new development tasks, and discuss next steps. 

\subsection{Overcoming the Terminology Gap}
Computer science and domain science use different jargon. This issue may complicate the understanding of the problem domain and project requirements since domain scientists tend to use rich vocabulary. The key to successful collaboration is keeping the vocabulary to the essentials and investing time in explaining new terms. Using many examples and frequent feedback from both sides will help bridge this gap.

At first, in this project, it was difficult to understand the domain requirements, due to terminology gaps. While many terms were familiar to domain scientists, getting a clear understanding of their exact meaning was challenging for computer scientists. As a result, there were numerous questions asked by both parties so as to understand the terminologies. Detailed written explanation of physics terminology with examples were provided by physicists and discussed during the meetings, so they can be used for future references as well as by new team members. 

Additionally, the development team requested that the domain scientists provide multiple examples of the functionalities expected from the portal.
Prototypes were used to arrive at a common understanding.

\subsection{Understanding the Domain Problem and Developing a Project Plan}
Xperts must devote substantial time to understand the domain problem, turning possibly vague ideas into a feasible approach, and developing a clear project plan.  At the start of the development of Version 2, a detailed document describing new functionalities with examples was developed by the combined team and thoroughly discussed to guide future work.  Having a clear understanding of the requirements for each version of the project is crucial and impacts many of the decisions that should be made along the way. Project requirements need to be well thought out, balanced, and clearly understood by all involved. It is important that the identified requirements are not dropped or compromised halfway through the project.

In this project, the requirements of different versions of the project were identified, documented, discussed, and sometimes prototyped to make sure they were feasible to be implemented and reflected what the team had envisioned and then implemented. The implementation was subjected to multiple revisions to ensure that it met the requirements of not only the physicists but also of the user community. Many examples were provided by both parties to ensure that the requirements were intelligible to team members. Prototyping helped team members visualize how the final pages would look like. It also helped in highlighting unanticipated design or technical challenges. By giving everyone a clear vision of the project, the prototype made everyone more involved in the process of building the pages.

\subsection{Prioritize Functional Requirements}

A dilemma that is often encountered in the process of planning any project is the existence of many desirable features but only a short project duration. Essential features should be differentiated from desirable requirements and should be strictly prioritized. Keep in mind that the requirements can change as new insights into a research project emerge, so re-assessments and re-prioritizing the requirements are necessary.

As part of this project, there was a discussion on how users can interact with the portal. One option was to display a periodic table to the user, with elements for which data were available being active and others greyed out. We decided, however, to prioritize elements and ions for which data was ready. We began with a simpler visual design of those elements only and created pages for each of their available properties.

\subsection{Source Code Management and Version Control}
Source code management and version control systems are used to track the evolution of an application. 
Source code management tools greatly facilitate collaborative software development. Such tools enable software roll-back to a previous, well-defined state and help developers start a new branch of the software. The latter can be important if two collaborators want to add their own separate feature sets.

Because we only wanted team members to have access to the repository, we used GitHub's private repository as the version control tool.

\subsection{Documentation}
There are many advantages to a well-documented application. Here, we will enlist those advantages that apply specifically to our project. Well-documented programs are easier to maintain and extend. Documentation makes it easier for new developers to reuse and extend code, which is a great concern for research teams where students graduate, leave, and are replaced by new students. Furthermore, documentation substantially enhances reproducibility.

For this purpose, we published documentation to be used exclusively by the project team on GitHub Pages. Along with Github repositories of project files, the use of GitHub pages to share instructions for generating the web pages greatly improved collaboration between team members. To evaluate the usefulness of the documentation to new users, testers from departments other than computer science were added to the project and were asked to follow instructions provided in the documentation to accomplish specific tasks. Their feedback helped us improve the documentation.

\subsection{ Maintainability \& Sustainability}
Sustainability~\cite{sustainability} ensures that the software will continue to be available in the future, on new platforms, meeting new needs. A requirement of this project was maintaining the portal for a long period of time without significant additional web development, even if funding were to be discontinued.
It is important for physicists to have the ability to add new elements and properties to the portal without extending the code. 

To fulfill this requirement, the process of creating web pages had to be automated. To this end, we standardized the file naming conventions and file content formats that were provided by the physics collaborators. A script to create templates for different properties was developed. Using these templates, HTML files were generated based on the contents of specific property files. A Python script can process and modify the files that are read in and generate HTML files or even graphical representations. 

The script has the ability to run on all elements and property files associated with those elements. This feature is added in case the template itself is changed, and we want all web pages created for that property to have the same look.
In addition, the script allows the user to pass the name of an individual element for which the property pages should be created. 
This option is used to add a new element or to update a specific data file.

Creating web pages is now as simple as dragging and dropping new data files into the repository and then running the script in one of its two available forms in order to generate the relevant web pages. 

\subsection{User Community Engagement and Exchange }

It is vital to engage the user community in the design and development of the user interface. Prototypes help obtain better feedback from the user community.

Giving presentations and talks on initial versions of this portal was a way to collect feedback. Moving forward, we will hold workshops to further engage the user community.

\section{Conclusion}
\label{sec:Conclusion}

The Atom Portal provides the user community with high-quality atomic data. 
The current version  offers access to the properties of 25 atoms and ions.

We have described the design challenges in creating the portal as well as 
the practices that have been applied in the development process. The practices have led to considerable time  savings in generating and updating the Atom Portal as well as in enhancing maintainability and extensibility.

Future releases will cover up to 100 atoms and ions. The implementation will be capable of retrieving data from a database as opposed to the current implementation of embedding the data directly into the web pages. Using dynamic web development, the creation of portal pages will be fully automated. Additionally, future updates will include the release of atomic computation codes, allowing researchers to reproduce some of the data currently provided through the portal.

\section{ACKNOWLEDGMENT}

This work was supported in part by the National Science Foundation under Awards No. OAC-1931339 and OAC-2209639. Research at Perimeter Institute is supported in part by the Government of Canada through the Department of Innovation, Science and Economic Development and by the Province of Ontario through the Ministry of Colleges and Universities. We thank Adams Marrs for work on Version~1 of the portal.

\begin{IEEEbiography}{Parinaz Barakhshan}{\,}is a Ph.D. candidate and Research Assistant at the University of Delaware, Department of Electrical and Computer Engineering. Her research interests include optimizing compilers and performance evaluation for High-Performance Computing systems. Contact her at parinazb@udel.edu.
\end{IEEEbiography}

\begin{IEEEbiography}{Akshay Bhosale}
is a Ph.D. candidate and Research Assistant in the Department of Electrical and Computer Engineering, University of Delaware. His research interests include Compilers, Optimization techniques and Automatic Parallelization. Contact him at akshay@udel.edu.
\end{IEEEbiography}

\begin{IEEEbiography}{Amani Kiruga}{\,} is an undergraduate researcher at the University of Delaware, Department of Computer and Information Sciences. His research interests include machine learning and computer vision. Contact him at akiruga@udel.edu.
\end{IEEEbiography}

\begin{IEEEbiography}{Rudolf Eigenmann}{\,} is a Professor at the Department of Electrical and Computer Engineering, University of Delaware, USA. His research interests include  optimizing compilers, programming methodologies, tools, and performance evaluation for high-performance computing, as well as the design of cyberinfrastructure. Contact him at eigenman@udel.edu.

\end{IEEEbiography}
\begin{IEEEbiography}{Marianna Safronova}{\,}is a Professor at the Department of Physics and Astronomy, University of Delaware, USA. Dr. Safronova's diverse research interests include atomic theory method and code development, quantum clocks, searches for new physics with quantum sensors, and others.   Contact her at msafrono@udel.edu.

\end{IEEEbiography}
\begin{IEEEbiography}{Bindiya Arora}{\,}is currently a PSI fellow at the Perimeter Institute for Theoretical Physics, Waterloo Canada. She is working as an Assistant Professor in the Department of Physics, Guru Nanak Dev University, India. Dr. Arora's research interests include high-precision atomic structure calculations for their applications in future technological developments.  Contact her at barora@perimeterinstitute.ca.
\end{IEEEbiography}

\end{document}